\documentclass{iaus}
\usepackage{graphicx}

\newcommand{\Msun}{\mathrm{M}_{\odot}}
\newcommand{\feh}{\mathrm{[Fe/H]}}
\def\apj{\textit{ApJ}}
\def\apjl{\textit{ApJ}}
\def\aj{\textit{AJ}}
\def\mnras{\textit{MNRAS}}
\def\araa{\textit{ARA\&A}}

\title[IAUS266.~~Globular Cluster Bimodality]{On the origin of globular cluster bimodality}

\author[Oleg Y. Gnedin]{Oleg Y. Gnedin}

\affiliation{University of Michigan, Department of Astronomy,
    Ann Arbor, MI 48109, USA \\ {\tt ognedin@umich.edu}}

\pubyear{2010}
\volume{266}
\jname{Star clusters: basic galactic building blocks}
\editors{R. de Grijs \& J. R. D. L\'epine, eds.}
\begin{document}

\maketitle

\begin{abstract}
Globular cluster systems in most large galaxies display bimodal color
and metallicity distributions, which are frequently interpreted as
indicating two distinct modes of cluster formation.  The metal-rich
(red) and metal-poor (blue) clusters have systematically different
locations and kinematics in their host galaxies.  However, the red and
blue clusters have similar internal properties, such as the masses,
sizes, and ages.  It is therefore interesting to explore whether both
metal-rich and metal-poor clusters could form by a common mechanism
and still be consistent with the bimodal distribution.  We show that
if all globular clusters form only during mergers of massive gas-rich
protogalactic disks, their metallicity distribution could be
statistically consistent with that of the Galactic globulars.  We take
galaxy assembly history from cosmological dark matter simulations and
couple it with the observed scaling relations for the amount of cold
gas available for star formation.  In the best-fit model, early
mergers of smaller hosts create exclusively blue clusters, whereas
subsequent mergers of progenitor galaxies with a range of masses
create both red and blue clusters.  Thus bimodality arises naturally
as the result of a small number of late massive merger events.  We
calculate the cluster mass loss, including the effects of two-body
scattering and stellar evolution, and find that more blue
clusters than red clusters are disrupted by the present time, because
of their smaller initial masses and larger ages.  The present-day mass
function in the best-fit model is consistent with the Galactic
distribution.  However, the spatial distribution of model clusters is
much more extended than observed and is independent of the parameters
of our model.  
\end{abstract}

\firstsection
\section{Are the Red and Blue Clusters Really That Different?}

A self-consistent description of the formation of globular clusters
remains a challenge to theorists.  A particularly puzzling observation
is the apparent bimodality, or even multimodality, of the color
distribution of globular cluster systems in galaxies ranging from
dwarf disks to giant ellipticals.  This color bimodality likely
translates into a bimodal distribution of the abundances of heavy
elements such as iron.  We know this to be the case in the Galaxy as
well as in M31, where relatively accurate spectral measurements exist
for a large fraction of the clusters.  The two most frequently
encountered modes are commonly called {\it blue} (metal-poor) and {\it
red} (metal-rich).

Bimodality in the globular cluster metallicity distribution of
luminous elliptical galaxies was proposed by \cite{zepf_ashman93},
following a theoretical model of \cite{ashman_zepf92}.  The concept of
cluster bimodality became universally accepted because the two
populations also differ in other observed characteristics.  The system
of red clusters have a significant rotation velocity similar to the
disk stars whereas blue clusters have little rotational support, in
the three disk galaxies observed in detail: Milky Way, M31, and M33.
In elliptical galaxies, blue clusters have a higher velocity
dispersion than red clusters, both due to lack of rotation and more
extended spatial distribution.  Red clusters are usually more
spatially concentrated than blue clusters (\cite{brodie_strader06}).
All of these differences, however, are in external properties
(location and kinematics), which reflect {\it where} the clusters
formed, but not {\it how}.  The internal properties of the red and
blue clusters are similar: masses, sizes, and ages, with only slight
differences.  Even the metallicities themselves differ typically by a
factor of 10 between the two modes, not enough to affect the dynamics
of molecular clouds from which these clusters formed.  Could it be
then that both red and blue clusters form in a similar way on small
scales, such as in giant molecular clouds, while the differences in
their metallicity and spatial distribution reflect when and where such
clouds assemble?

All scenarios proposed in the literature assumed different formation
mechanisms for the red and blue clusters, and most scenarios
envisioned the stellar population of one mode to be tightly linked to
that of the host galaxy (e.g., \cite{forbes_etal97},
\cite{cote_etal98}, Strader et al.~2005).  The other mode is assumed
to have formed differently, in some unspecified ``primordial'' way.
This assumption only pushed the problem back in time but it did not
solve it.  For example, \cite{beasley_etal02} used a semi-analytical
model of galaxy formation to study bimodality in luminous elliptical
galaxies and needed two separate prescriptions for the blue and red
clusters.  In their model, red clusters formed in gas-rich mergers
with a fixed efficiency of 0.007 relative to field stars, while blue
clusters formed in quiescent disks with a different efficiency of
0.002.  The formation of blue clusters also had to be artificially
truncated at $z=5$.  \cite{strader_etal05} and \cite{rhode_etal05}
suggested that the blue clusters could instead have formed in very
small halos at $z > 10$, before cosmic reionization removed cold gas
from such halos.  This scenario requires high efficiency of cluster
formation in the small halos and also places stringent constraints on
the age spread of blue clusters to be less than 0.5 Gyr.
Unfortunately, even the most recent measurements of the relative
cluster ages in the Galaxy (\cite{deangeli_etal05,
marin-franch_etal09}) cannot detect age differences smaller than 9\%,
or about 1 Gyr, and therefore cannot support or falsify the
reionization scenario.

\section{Globular Clusters Could Form in Protogalactic Disks}

We set out to test whether a common mechanism could explain the
formation of both red and blue modes and produce an entire metallicity
distribution consistent with the observations.  We begin with a
premise of the hierarchical galaxy formation in a $\Lambda$CDM
universe.  Primordial density fluctuations in the early universe,
probed directly by the CMB anisotropies, set the seeds for structure
formation.  Cosmological numerical simulations study the growth of the
fluctuations via gravitational instability and show us a history of
galaxy assembly.  The simulations begin with tiny deviations from the
Hubble flow, whose amplitudes are set by the measured power spectrum
while the phases are assigned randomly.  Therefore, each particular
simulation provides only a statistical description of a representative
part of the Universe, although current models successfully reproduce
major features of observed galaxies.

\begin{figure}[t]
\centering
\vspace{0.2cm}
\includegraphics[width=2.47in]{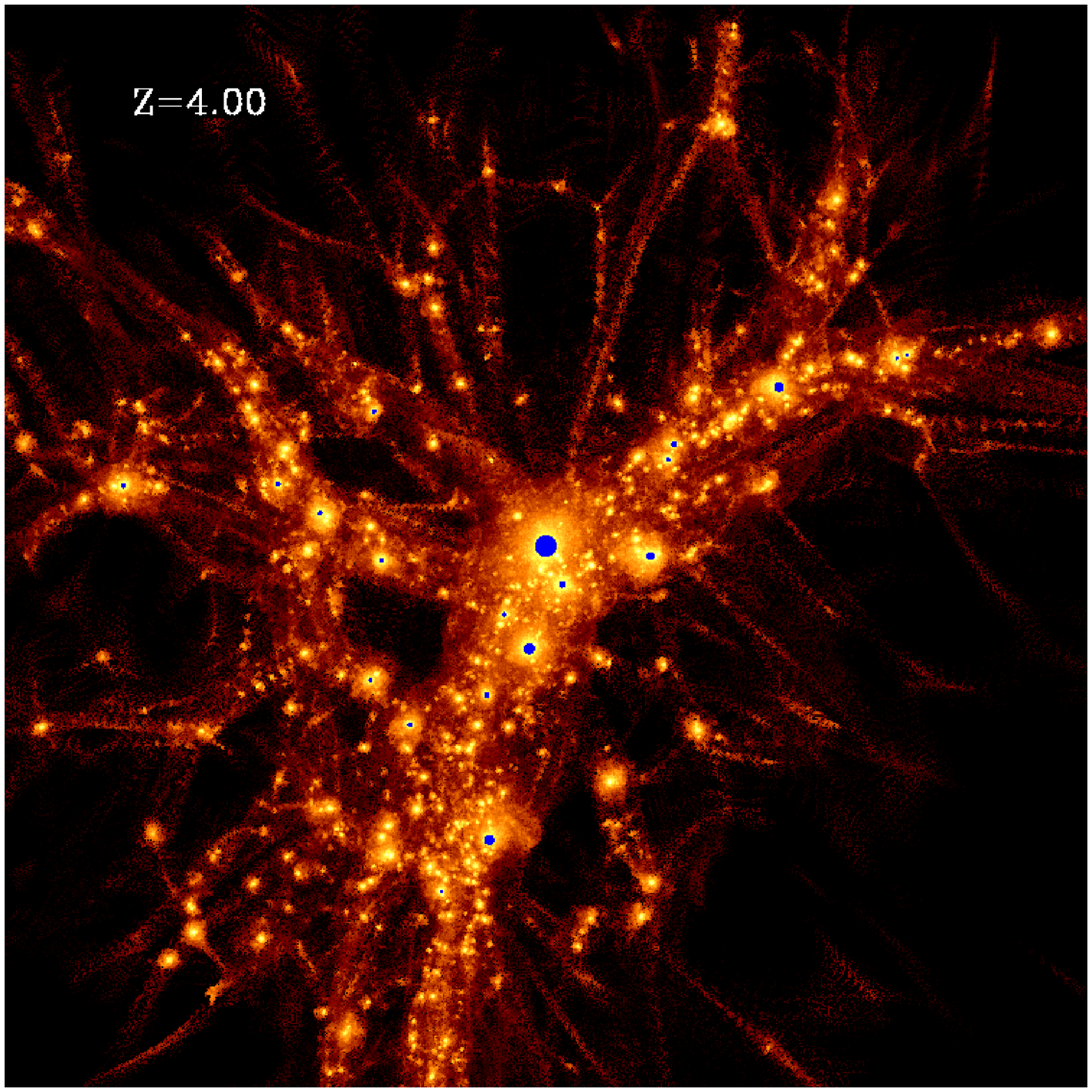}
\includegraphics[width=2.8in]{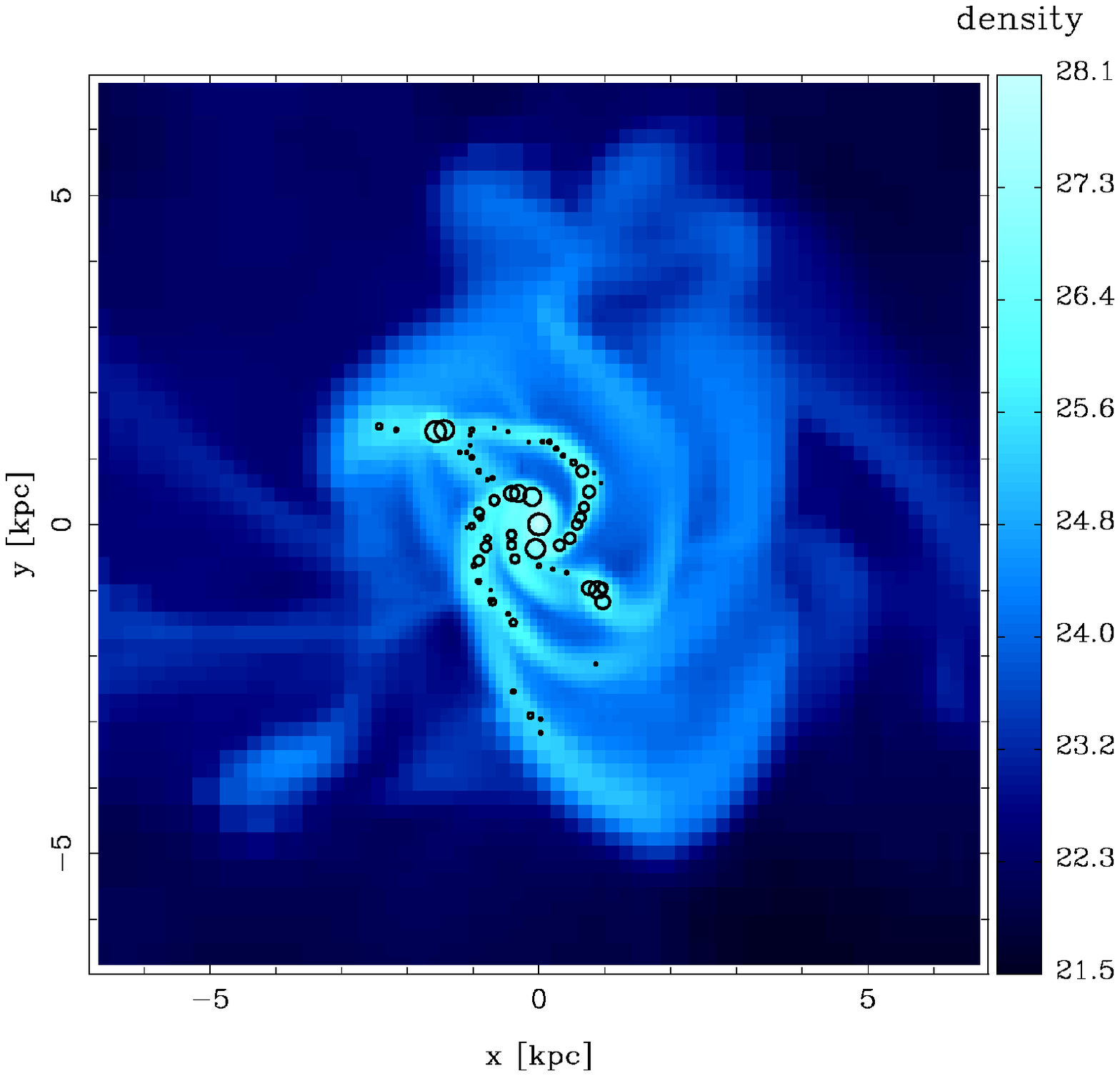}
\caption{{\it Left:} Hierarchical build-up of a Milky Way-type galaxy
  in an adaptive mesh simulation at $z=4$. The view is centered on the
  largest progenitor galaxy in the simulation and shows $1 \, h^{-1}$
  Mpc region (comoving).  Dark circles show protogalactic disks in the
  central regions of dark matter halos. {\it Right:} A massive gaseous
  disk with prominent spiral arms, seen face-on, in the process of
  active merging at $z=4$.  In Kravtsov \& Gnedin (2005) model star
  clusters form in giant gas clouds, shown by circles with the sizes
  corresponding to the cluster masses.}
\vspace{0.2cm}
  \label{fig:gc}
\end{figure}

Hubble Space Telescope observations have convincingly demonstrated one
of the likely formation routes for massive star clusters today -- in
the mergers of gas-rich galaxies (e.g., \cite{holtzman_etal92,
oconnell_etal95, whitmore_etal99, zepf_etal99}).  We adopt this single
formation mechanism and assume that globular clusters form only during
massive gas-rich mergers.  We follow the merging process of progenitor
galaxies in a Galaxy-sized environment using a set of cosmological
$N$-body simulations from \cite{kravtsov_etal04}.  We need to decide
what type and how many clusters will form in each merger event.  For
this purpose, we use observed scaling relations to assign each dark
matter subhalo a certain amount of cold gas that will be available for
star formation throughout cosmic time and an average metallicity of
that gas.  In order to keep the model transparent, we choose as simple
a parametrization of the cold gas mass as possible.  Finally, we make
the simplest assumption that the mass of all globular clusters formed
in the merger is linearly proportional to the mass of this cold gas,
$M_{GC} \propto M_g$.  We discuss our results in \S\ref{sec:met}.

Although such model appears extremely simplistic, we have some
confidence that it may capture main elements of the formation of
massive clusters.  \cite{kravtsov_gnedin05} used a hydrodynamic
simulation of the Galactic environment at high redshifts $z > 3$ and
found dense, massive gas clouds within the protogalactic clumps.
These clouds assemble during gas-rich mergers of progenitor galaxies,
when the cold gas forms a thin, self-gravitating disk.  The disk
develops strong spiral arms, which further fragment into separate
molecular clouds located along the arms as beads on a string (see
Fig. \ref{fig:gc}).  If the high-density regions of these clouds
formed star clusters, the resulting distributions of cluster mass,
size, and metallicity are consistent with those of the Galactic
metal-poor clusters.  The high stellar density of Galactic clusters
restricts their parent clouds to be in relatively massive progenitors,
with the total mass $M_h > 10^9\ \Msun$.  The mass of the molecular
clouds increases with cosmic time, but the rate of mergers declines
steadily.  Therefore, the cluster formation efficiency peaks during an
extended epoch, $5 < z < 3$, when the Universe is less than 2 Gyr old.
The molecular clouds are massive enough to be shielded from the
extragalactic UV radiation, so that globular cluster formation is
unaffected by the reionization of cosmic hydrogen.  The mass function
of model clusters is consistent with a power law $dN/dM \propto
M^{-\alpha}$, where $\alpha = 2.0 \pm 0.1$, similar to the local young
star clusters.  The total mass of clusters formed in each progenitor
is roughly proportional to the available gas supply and the total
mass, $M_{GC} \propto M_g \propto M_h$.

\begin{figure}[t]
\centering
\includegraphics[height=2.4in]{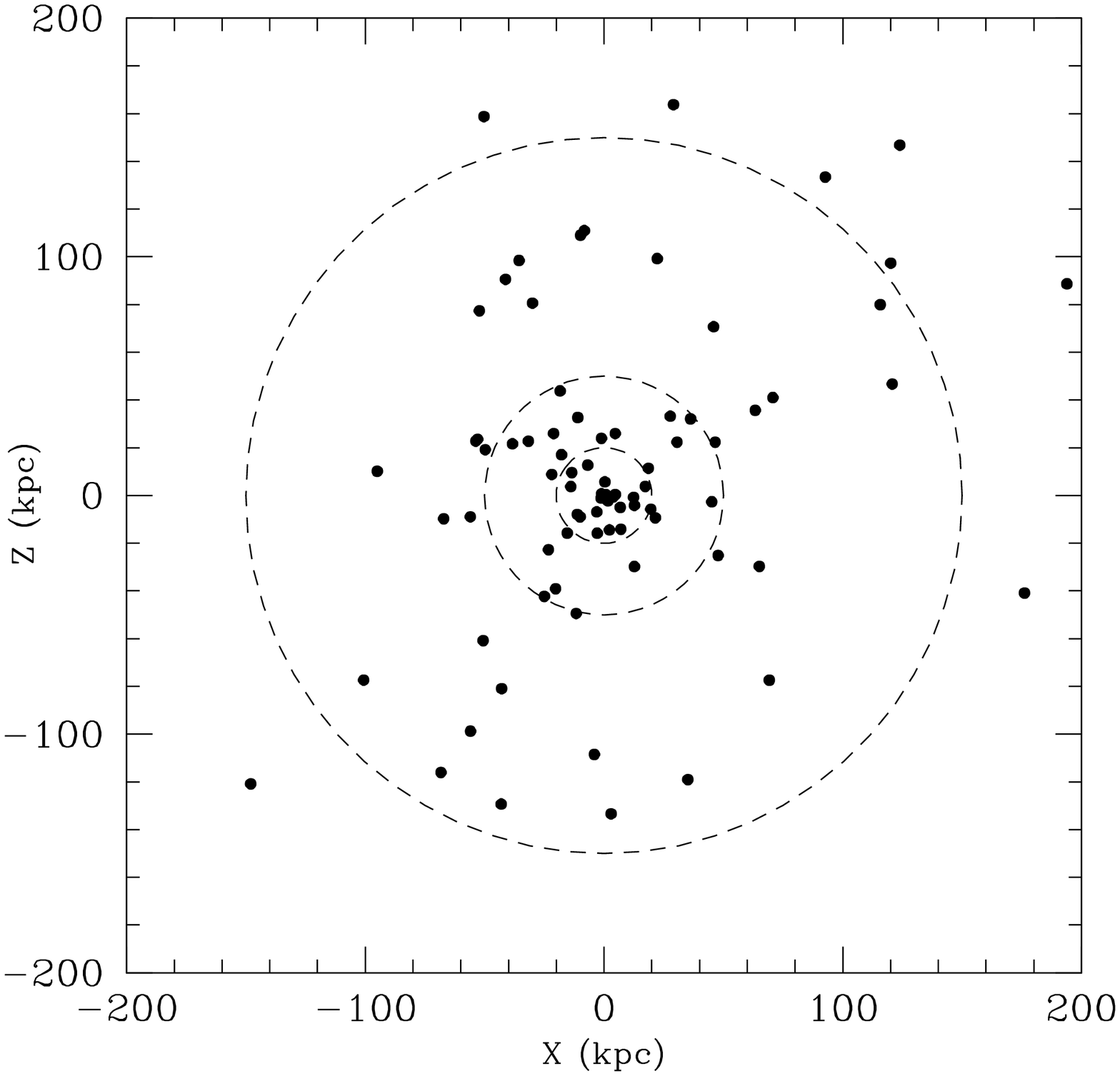}
\includegraphics[height=2.4in]{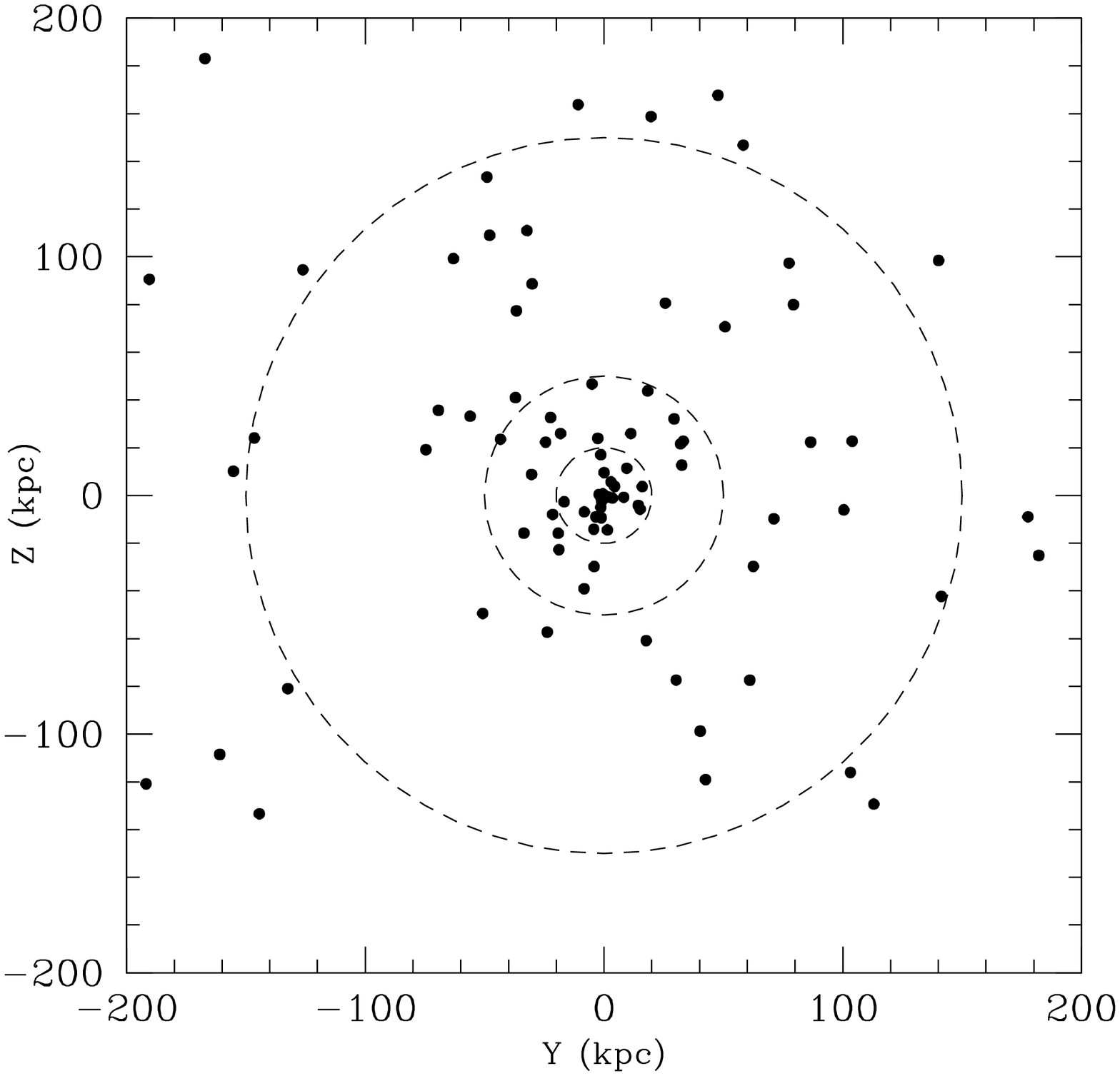}
\includegraphics[height=2.4in]{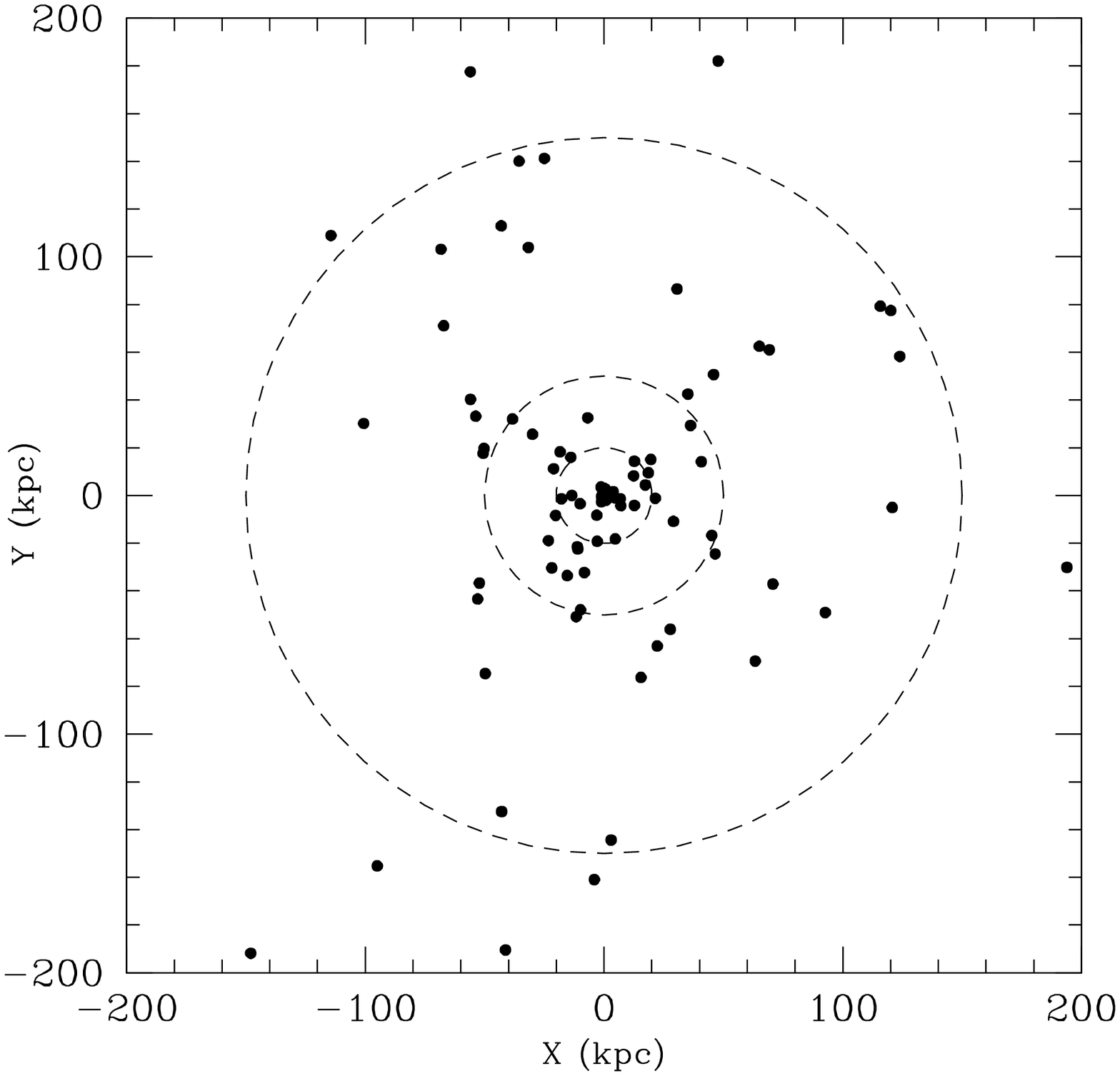}
\includegraphics[height=2.4in]{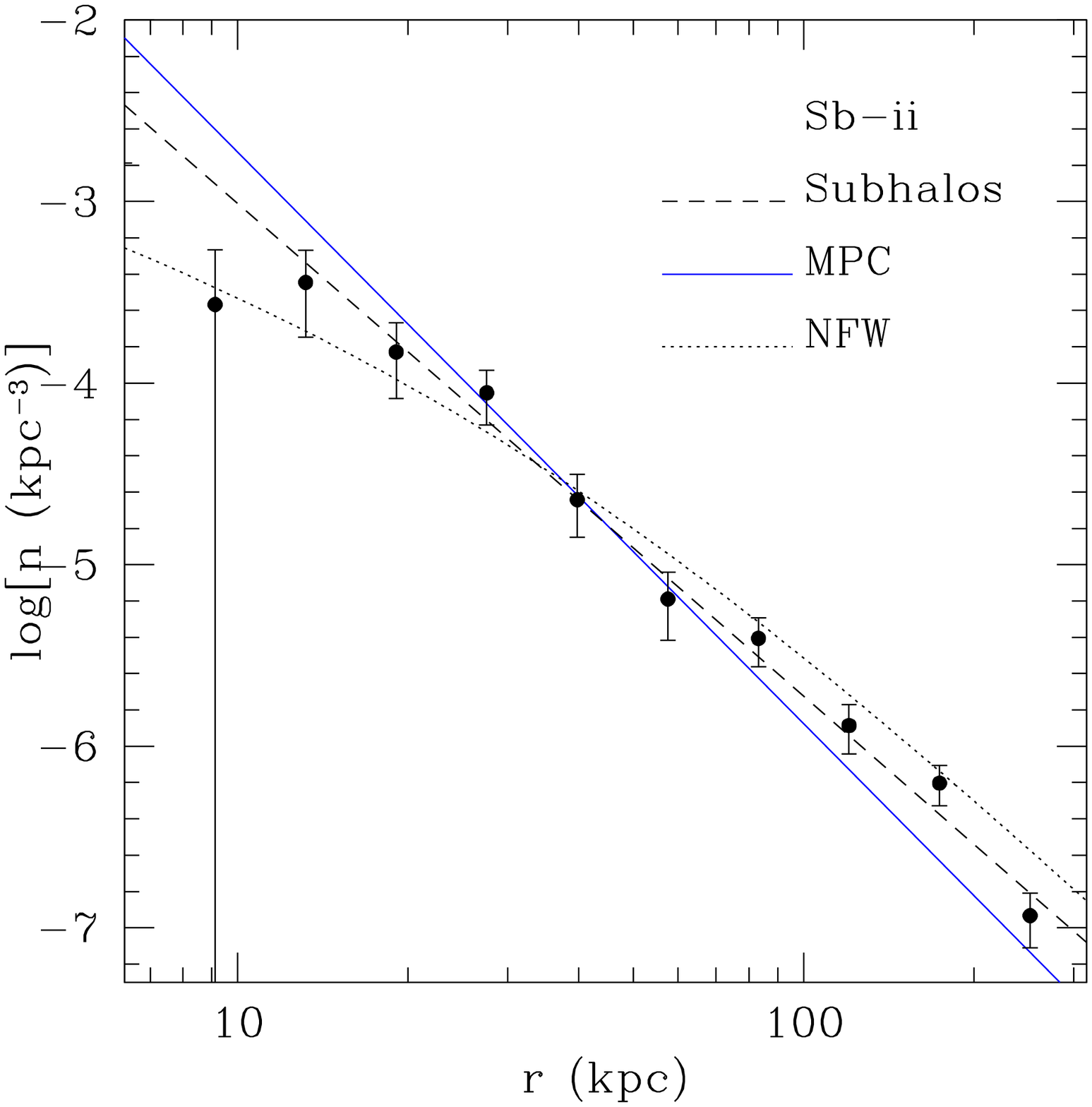}
\caption{Spatial distribution of surviving model globular clusters,
  calculated by Prieto \& Gnedin (2008).  Dashed circles are to
  illustrate the projected radii of 20, 50, and 150 kpc.  The number
  density profile ({\it bottom right}) can be fit by a power law,
  $n(r)\propto r^{-2.7}$.  The distribution of model clusters is
  similar to that of surviving satellite halos ({\it dashed line}) and
  smooth dark matter ({\it dotted line}).  It is also consistent with
  the observed slope of the metal-poor globular clusters in the
  Galaxy ({\it solid line}), plotted using data from the catalog of
  Harris (1996).}
  \label{fig:density}
\end{figure}

In \cite{prieto_gnedin08} we showed that subsequent mergers of the
progenitor galaxies ensure that the present distribution of the
globular cluster system is spheroidal, as observed.  Since the
hydrodynamic simulation was stopped at $z \approx 3.3$, we used the
Kravtsov et al. (2004) $N$-body simulation in order to calculate
cluster orbits to $z=0$.  We used the evolving properties of all
progenitor halos, from the outputs with a time resolution of $\sim
10^8$ yr, to derive the gravitational potential in the whole
computational volume at all epochs.  We calculated the orbits of
globular clusters in this potential from the time when their host
galaxies accrete onto the main (most massive) galaxy.  Using these
orbits, we calculated the dynamical evolution of model clusters,
including the effects of stellar mass loss, two-body relaxation, tidal
truncation, and tidal shocks.

In this model, all clusters form on nearly circular orbits within the
protogalactic disks.  Depending on the subsequent trajectories of the
hosts, clusters form three main subsystems at present time.  {\it Disk
clusters} formed in the most massive progenitor that eventually hosts
the present Galactic disk.  These clusters, found within the inner 10
kpc, do not actually stay on circular orbits but instead are scattered
to eccentric orbits by perturbations from accreted galactic
satellites.  {\it Inner halo clusters}, found between 10 and 60 kpc,
came from the now-disrupted satellite galaxies.  Their orbits are
inclined with respect to the Galactic disk and are fairly isotropic.
{\it Outer halo clusters}, beyond 60 kpc from the center, are either
still associated with the surviving satellite galaxies, or were
scattered away from their hosts during close encounters with other
satellites and consequently appear isolated.

The azimuthally-averaged space density of globular clusters is
consistent with a power law, $n(r)\propto r^{-\gamma}$, with the slope
$\gamma \approx 2.7$ (see Fig. \ref{fig:density}).  Since all of the
distant clusters originate in progenitor galaxies and share similar
orbits with their hosts, the distribution of the clusters is almost
identical to that of the surviving satellite halos.  This power law is
similar to the observed slope of the metal-poor ($\feh < -1$)
globular clusters in the Galaxy.  However, the model clusters have a
more extended spatial distribution (larger median distance) than
observed.  In the model it is largely determined by the orbits of the
progenitor galaxies and the epoch of formation.  \cite{moore_etal06}
showed that the early-forming halos are more spatially concentrated
and in order to match the Galactic distribution, globular clusters
would need to form at $z \sim 12$.  However, such an early formation
is inconsistent with the requirement of high mass and density of the
parent molecular clouds.  At present, we do not have an acceptable
solution to this problem.

\begin{figure}[t]
\centering
 \includegraphics[width=2.63in]{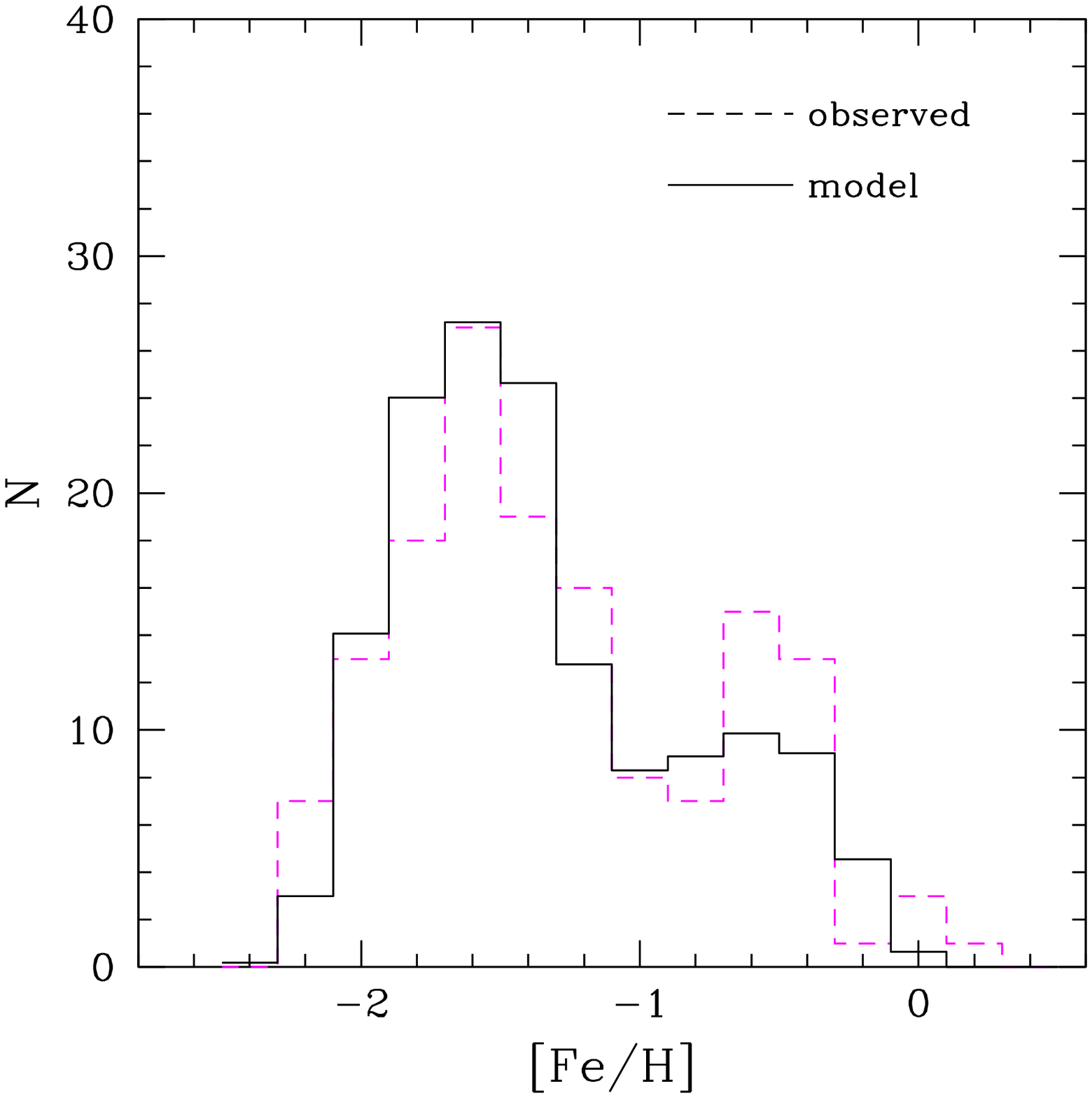}
 \includegraphics[width=2.63in]{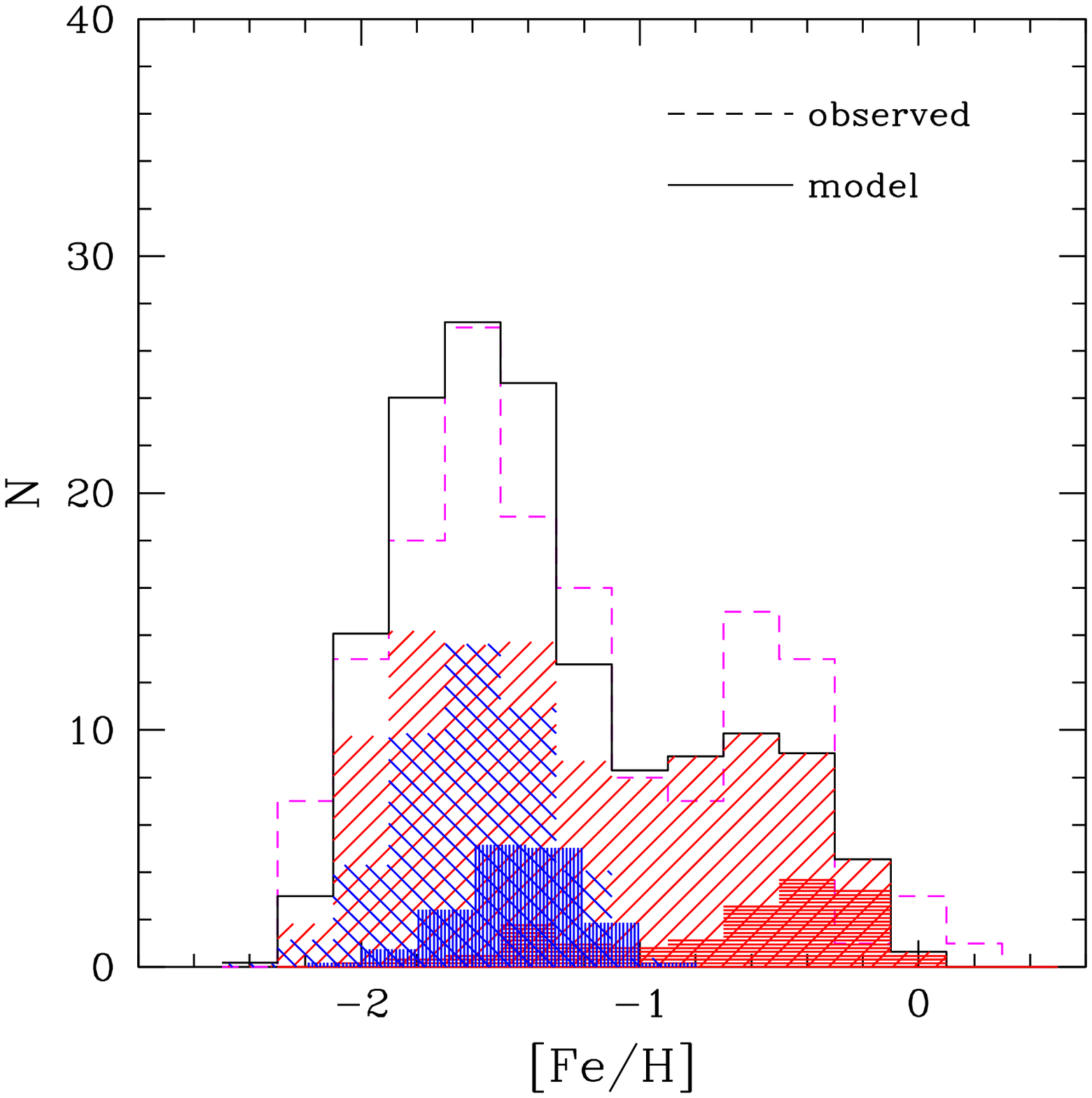}
\vspace{-0.6cm}
\caption{Metallicities of model clusters at $z=0$ (solid histogram)
  compared to the observed distribution of Galactic globular clusters
  (shaded histogram).  {\it Right panel:} Contributions of major
  mergers (red) and early mergers (blue) to the total model count.
  Filled histograms are for the main Galactic disk.  
  From Muratov \& Gnedin, in prep.}
\vspace{0.4cm}
  \label{fig:gc_metal}
\end{figure}

\section{Metallicity Bimodality as a Natural Outcome of Hierarchical Galaxy Formation}
  \label{sec:met}

Following the scenario outlined above, A. Muratov \& O. Gnedin (in
prep.) developed a semi-analytical model that aims to reproduce
statistically the metallicity distribution of the Galactic globular
clusters, as compiled by \cite{harris96}.  The formation of clusters
is triggered during a merger of gas-rich protogalaxies with the mass
ratio 1:5 or higher, and during very early mergers with any mass ratio
when the cold gas fraction in the progenitors is close to 100\%.
These criteria are applied at every timestep of the simulation, every
$\sim 10^8$ yr.  

In the best-fit model, the mass of globular clusters
formed in each event is
$$
  M_{GC} \approx 7\times 10^{-4}\ M_g \approx 10^{-4}\ M_h.
$$ 
This imposes the minimum mass of a halo capable of forming a
globular cluster.  Based on dynamical disruption arguments we track
only the clusters more massive than $M_{\rm min} = 10^5\, \Msun$.
Therefore, to form even a single cluster with the minimum mass, the
halo needs to be more massive than $\sim 10^9\, \Msun$.  Individual
cluster masses are drawn randomly from the assumed initial cluster
mass function, $dN/dM \propto M^{-2}$, normalized to $M_{GC}$.  The
clusters are assigned the mean metallicity $\feh$ of their host
protogalaxies, using the observed galaxy stellar mass-metallicity
relation.  Overall the model has 5 free parameters relating to the
normalization of the cluster formation rate and the merger mass ratio.

Figure \ref{fig:gc_metal} shows the metallicity distribution in the
best-fit model.  The red peak is not as pronounced as in the
observations but is noticeable.  Interestingly, major mergers
contribute both to the red and blue modes, in about equal proportions.
Early mergers of low-mass progenitors contribute only blue clusters.

Note that the model has the same formation criteria for all clusters,
without explicitly differentiating between the two modes.  The only
variables are the gradually changing amount of cold gas, the growth of
protogalactic disks, and the rate of merging.  Yet, the model produces
two peaks of the metallicity distribution, centered at $\feh \approx
-1.6$ and $\feh \approx -0.6$, matching the Galactic globular
clusters.  The KS probability of the model being consistent with the data
is $P_{KS} = 80\%$.

\begin{figure}[t]
\centering
 \includegraphics[width=3.5in]{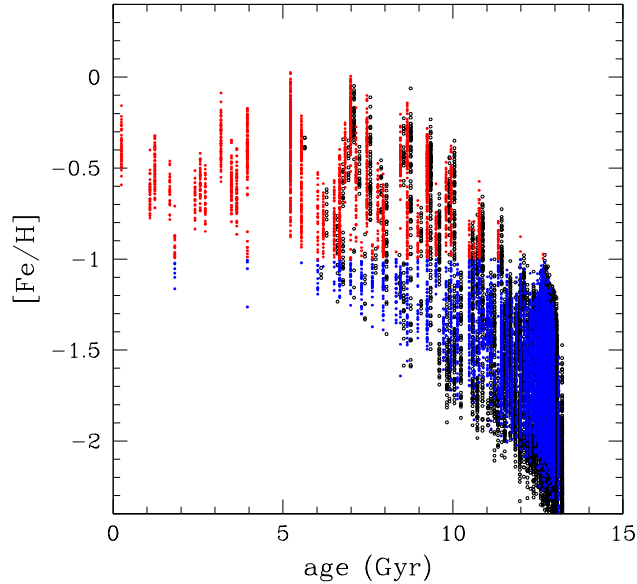}
\vspace{-0.4cm}
\caption{Age-Metallicity relation of model clusters in 64 realizations
  of the Galactic sample.  Filled (colored) circles show surviving
  clusters, open (black) circles show disrupted clusters.  The
  build-up of massive halos drives the steep slope of this relation at
  early epochs.  Notice an order-of-magnitude spread in metallicities
  of clusters forming at a given epoch.}
\vspace{0.2cm}
  \label{fig:age_metal}
\end{figure}

Our prescription links cluster metallicity to the average galaxy
metallicity in a one-to-one relation, albeit with random scatter.
Since the average galaxy metallicity grows monotonically with time,
the cluster metallicity also grows with time.  The model thus encodes
an age-metallicity relation, in the sense that metal-rich clusters
are younger than their metal-poor counterparts by several Gyr.
However, Figure \ref{fig:age_metal} shows that clusters of the same
age may differ in metallicity by as much as a factor of 10, as they
formed in the progenitors of different mass.  Observations of the
Galactic globular clusters do not show a clear age-metallicity
relation, but instead indicate an age spread increasing with
metallicity (\cite{deangeli_etal05, marin-franch_etal09}).  Our model
does not appear to be in an obvious conflict with this trend.

\begin{figure}[t]
\centering
 \includegraphics[width=2.63in]{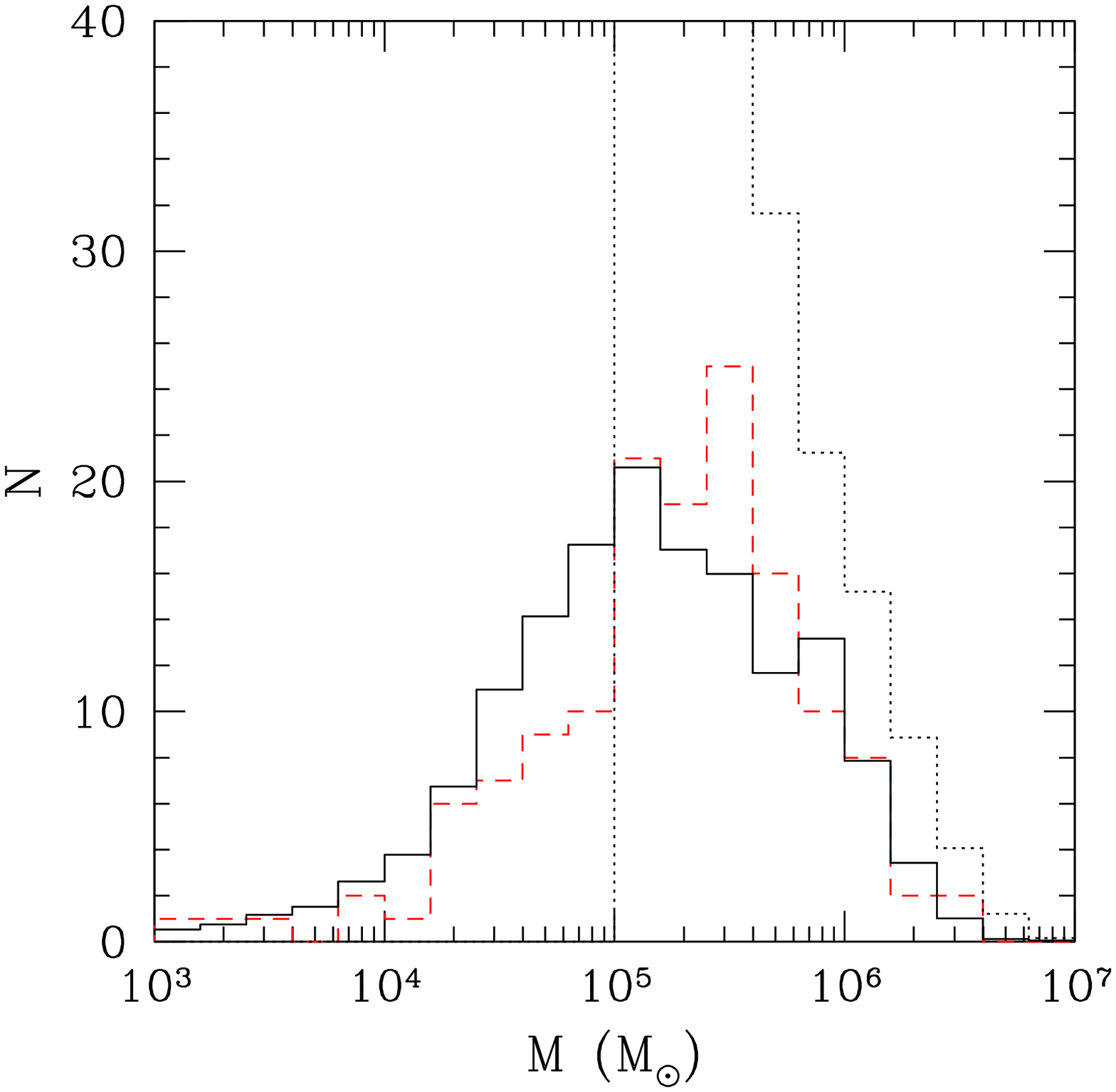}
 \includegraphics[width=2.63in]{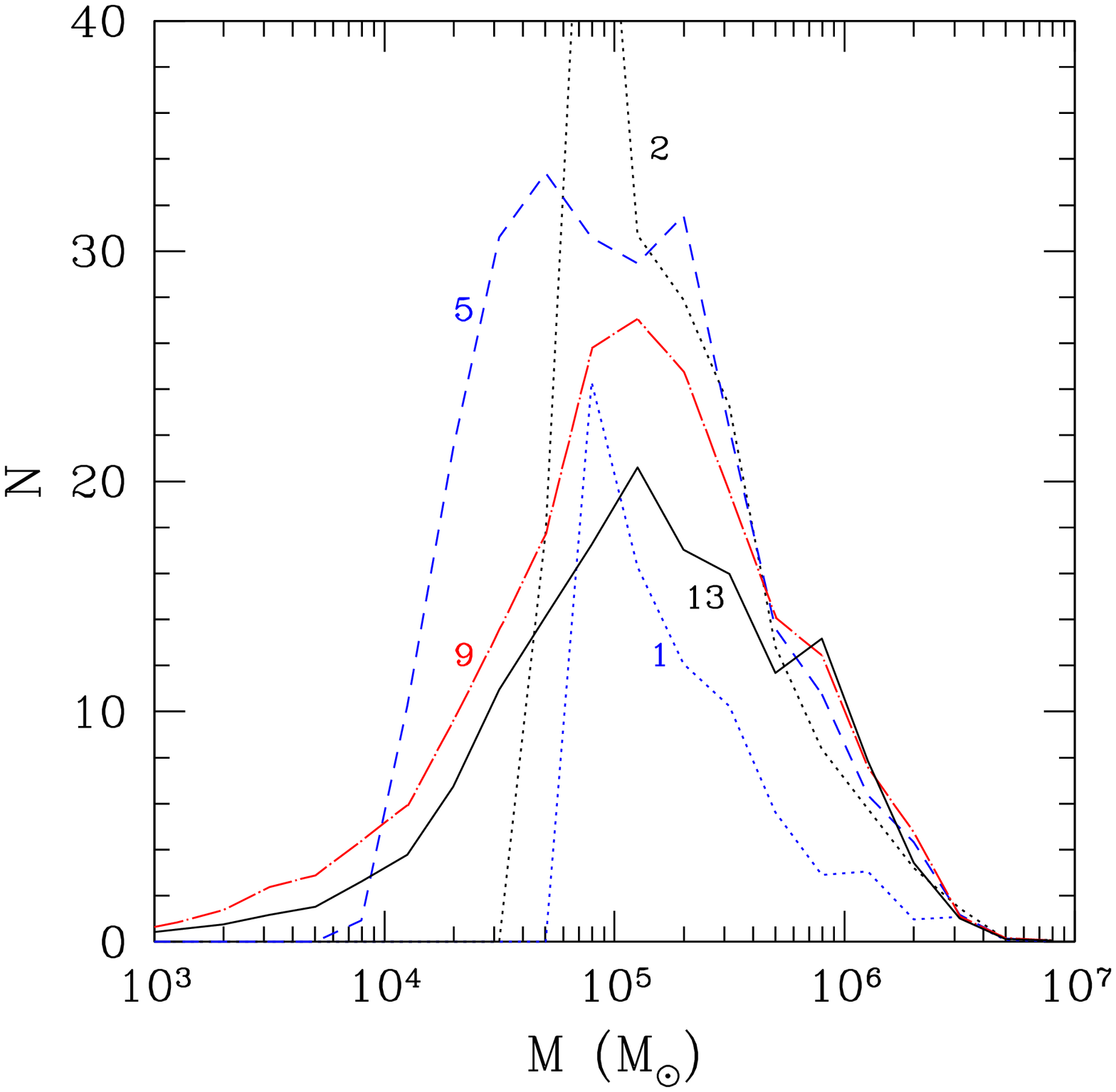}
\vspace{-0.6cm}
\caption{{\it Left:} Mass function of model clusters at $z=0$ (solid
  histogram) vs. Galactic clusters (dashed histogram).  Dotted
  histogram shows the combined initial masses of model clusters formed
  at all epochs, including those that did not survive until the
  present.  We do not follow clusters with the initial masses below
  $10^5\ \Msun$.  {\it Right:} Actual mass function at cosmic
  times of 1 Gyr ($z \approx 5.7$, dotted), 2 Gyr ($z \approx 3.2$,
  dotted), 5 Gyr ($z \approx 1.3$, dashed), 9 Gyr ($z \approx 0.5$,
  dot-dashed), and 13 Gyr ($z \approx 0$, solid).}
\vspace{0.4cm}
  \label{fig:mf}
\end{figure}

Some of the old and low-mass clusters will be disrupted by the gradual
escape of stars and will not appear in the observed sample.  We
calculated the effects of the dynamical evolution of model clusters,
including stellar mass loss and two-body evaporation, but ignored
tidal shocks for simplicity.  Figure \ref{fig:mf} compares the
resulting model mass function at $z=0$ with the observed Galactic
distribution.  Since the model parameters were tuned to reproduce the
metallicity distribution, the mass functions do not match as well but
are still consistent at the level of $P_{KS} = 7\%$.  Majority of the
disrupted clusters were blue clusters that formed in early low-mass
progenitors.  

Right panel of Figure \ref{fig:mf} illustrates the evolution of the
mass function as an interplay between the continuous buildup of
massive clusters ($M > 10^5\, \Msun$) and the dynamical erosion of the
low-mass clusters ($M < 10^5\, \Msun$).  Expecting that the clusters
below $M_{\rm min}$ would eventually be disrupted, we did not track
their formation in the model.  Instead, the low end of the mass
function was built by the gradual evaporation of more massive
clusters.  Note that most of the clusters were not formed until the
universe was 2 Gyr old, corresponding to $z \approx 3$.  The fraction
of clusters formed before $z \approx 6$, when cosmic hydrogen was
reionized, is small.

In calculating the rate of two-body relaxation, we assumed a standard
result for the evaporation time, $t_d(M) \approx 10^{10} (M/2\times
10^5\, \Msun)$ yr.  We have also used an alternative rate with the
weaker mass dependence, $t_d(M) \approx 10^{10} (M/2\times 10^5\,
\Msun)^{2/3}$ yr, suggested by the recent results of
\cite{gieles_baumgardt08}.  While this prescription leads to slower
disruption of the low-mass clusters, we find that the resulting
distribution at $z=0$ is still consistent with observations at the
level of $P_{KS}$ of a few percent.

\bigskip

In our scenario, bimodality results from the history of galaxy
assembly (rate of mergers) and the amount of cold gas in protogalactic
disks.  Early mergers are very frequent but involve relatively
low-mass protogalaxies, which produce preferentially blue clusters.
Late mergers are infrequent but typically involve more massive
galaxies.  As the number of clusters formed in each merger increases
with the progenitor mass, just a few late super-massive mergers can
produce a significant number of red clusters.  The concurrent growth
of the average metallicity of galaxies between the late mergers leads
to an apparent ``gap'' between the red and blue clusters.

We expect that our formalism could be applied to other galactic
environments, such as those of elliptical galaxies with larger samples
of globular clusters.  For example, \cite{peng_etal08} showed that the
fraction of red clusters increases from 10\% to 50\% with increasing
luminosity of elliptical galaxies in the Virgo cluster.  Compared to
the Galaxy, giant ellipticals are expected to experience more massive
mergers, which would produce similar numbers of red and blue clusters,
according to Figure \ref{fig:gc_metal}.  Thus a red cluster fraction
peaking at $\sim 50\%$ is a natural outcome of the hierarchical
formation.


\end{document}